\documentclass{nature}

\usepackage{graphicx}
\usepackage{amsmath,amssymb}
\usepackage{bm}
\usepackage{booktabs}
\usepackage{multirow}
\usepackage{url}
\usepackage{xcolor}
\usepackage{caption}
\usepackage{float}
\usepackage[colorlinks=true, linkcolor=blue, citecolor=blue, urlcolor=blue]{hyperref}

\DeclareCaptionFormat{bold}{\textbf{#1#2}#3}
\graphicspath{{Pic/}}

\makeatletter
\renewenvironment{figure}
   {\@float{figure}}
   {\end@float}
\makeatother

\title{Criticality and reduced dynamical resilience in PM$_{2.5}$ pollution systems}

\author{Yuan Chen$^{1,2,8}$, Yongwen Zhang$^{1,2,8,\ast}$, Xu Li$^{1,2}$, Dean Chen$^{3}$, Jingfang Fan$^{4}$, Yosef Ashkenazy$^{5}$, Deliang Chen$^{6}$, Shlomo Havlin$^{7}$}

\begin{document}
\maketitle

\begin{affiliations}
\item Yunnan Key Laboratory of Complex Systems and Brain-Inspired Intelligence, Kunming University of Science and Technology, Kunming 650500, Yunnan, China;
\item Faculty of Science, Kunming University of Science and Technology, Kunming 650500, Yunnan, China;
\item Institute for Atmospheric and Earth System Research/Physics, Faculty of Science, University of Helsinki, Helsinki, Finland;
\item School of Systems Science/Institute of Nonequilibrium Systems, Beijing Normal University, Beijing, China;
\item Department of Solar Energy and Environmental Physics, The Jacob Blaustein Institutes for Desert Research, Ben-Gurion University of the Negev, Midreshet Ben-Gurion 84990, Israel;
\item Department of Earth System Science, Tsinghua University, Beijing, 100084, China;
\item Department of Physics, Bar-Ilan University, Ramat Gan 52900, Israel.
\\ \normalsize{$^8$These authors contributed equally: Yuan Chen, Yongwen Zhang.}
\\ \normalsize{$^\ast$Correspondence and requests for materials should be addressed to Yongwen Zhang.}
\\ \normalsize{e-mail: zhangyw@kust.edu.cn}
\end{affiliations}

\vspace{2em}


\begin{abstract}
Concentration-based metrics underpin air-quality assessment, while dynamical persistence and recovery describe how rapidly high-PM$_{2.5}$ episodes dissipate and how strongly they retain memory. Here we introduce a finite-memory multiplicative reversion (FMMR) process that links the lognormal concentration backbone of PM$_{2.5}$ variability with event recurrence, temporal memory, variance amplification and local dynamical resilience. Across station observations and reanalysis data, elevated PM$_{2.5}$ regimes show a coherent set of critical signatures: stronger memory, rising autocorrelation, broader upper tails, amplified variance, reduced resilience and more clustered exceedance events. Together, these co-occurring signals reveal dynamical criticality in PM$_{2.5}$ pollution systems, with critical slowing down expressed as a loss of restoring capacity under high-pollution conditions. A gridded comparison across populated and emission-influenced regions further shows that areas with similar PM$_{2.5}$ burden can differ in recovery capacity, while eastern China has shifted toward higher resilience during recent air-quality improvements and India and West Africa occupy lower-resilience states. By identifying where pollution burden and recovery capacity diverge, these findings establish dynamical persistence and resilience as complementary dimensions of PM$_{2.5}$ risk and provide a quantitative basis for resilience-oriented air-quality assessment.

\end{abstract}

\vspace{2em}

\section*{INTRODUCTION}

Densely populated and emission-influenced regions concentrate people, infrastructure and economic activity, so persistent air-pollution episodes can amplify health, mobility and economic risks~\cite{Kim2025UrbanAirGlobal,Liang2020UrbanAir}. Among these pollutants, fine particulate matter (PM$_{2.5}$), particles with aerodynamic diameters smaller than 2.5 micrometers, represents one of the most severe threats to public health, contributing to approximately 4.2 million premature deaths worldwide in 2019~\cite{GBD2019PM25}. Beyond its direct health impacts, elevated PM$_{2.5}$ concentrations can disrupt transportation systems, reduce visibility, and constrain economic activity, thereby amplifying systemic risks in exposed urban and regional environments~\cite{Han2016PM25Visibility,Dai2017EconomicPM25}.

Air-pollution risk is commonly quantified using concentration-based metrics, including annual means, threshold exceedances and population-weighted exposure, which underpin health assessment and regulatory management~\cite{GBD2019PM25,Kim2025UrbanAirGlobal}. A complementary dynamical perspective asks how long high-pollution episodes persist, how rapidly concentrations recover after perturbations and how strongly present pollution levels depend on past conditions~\cite{Scheffer2009EWS,Dakos2012Methods}. Regions with comparable mean PM$_{2.5}$ levels may therefore differ substantially in dynamical stability and recovery capacity. These properties are central to pollution resilience, because persistent episodes can prolong exposure and amplify societal impacts even when average concentrations appear similar.

Criticality provides a powerful framework for interpreting persistence and resilience loss in complex systems. Near critical regimes, systems often exhibit slower recovery, enhanced memory, amplified fluctuations and clustered events, reflecting weaker effective restoring forces~\cite{Scheffer2009EWS,Dakos2012Methods}. In atmospheric pollution systems, these signatures are expected when pollutant accumulation, unfavorable meteorology and inefficient removal jointly increase residence time and temporal dependence. High PM$_{2.5}$ episodes are often associated with shallow and stable boundary layers, weak ventilation, stagnant synoptic conditions, humidity-dependent aerosol formation and reduced wet removal~\cite{8,9,10,11,12,13,Ju2022BLM}. We use criticality in this dynamical sense: a measurable loss of restoring capacity, growing persistence and stronger fluctuations. Testing this interpretation requires indicators that connect the distributional structure of PM$_{2.5}$ variability with event recurrence, temporal memory and recovery.

The concentration distribution provides a natural starting point for this diagnosis. Ambient PM$_{2.5}$ concentrations typically exhibit right-skewed distributions that are approximately lognormal, consistent with multiplicative processes governing pollutant accumulation, dispersion, transformation and removal~\cite{1,2}. Under elevated pollution conditions, these distributions broaden and their upper tails become progressively more extended~\cite{4,5}. Such behavior has been discussed in relation to critical dynamics and self-organized criticality in air-pollution systems~\cite{44,55,6,7}. Other dynamical indicators, including variance, lag-1 autocorrelation, recovery time and waiting-time statistics between pollution events, provide complementary information about persistence and resilience~\cite{Scheffer2009EWS,Dakos2012Methods}. This interpretation is also motivated by nonequilibrium multiplicative-noise systems, in which stochastic fluctuations and temporal correlations can generate phase-transition-like behavior~\cite{VanDenBroeck1994Noise,Mangioni2000MultiplicativeNoise}.

Despite these insights, air-quality assessment still lacks a unified stochastic framework that links multiplicative concentration variability, event recurrence, temporal memory and pollution resilience. Existing atmospheric interpretations emphasize emissions, boundary-layer meteorology, synoptic transport, aerosol chemistry and deposition, whereas statistical indicators of persistence and criticality are often examined separately. This separation makes it difficult to determine whether distributional scaling, clustered recurrence and slower recovery arise from a common dynamical structure, and limits the ability to distinguish regions where PM$_{2.5}$ burden is high from regions where recovery capacity is weak.

Here we introduce a Finite-Memory Multiplicative Reversion (FMMR) process to characterize the stochastic dynamics of atmospheric PM$_{2.5}$ concentrations and to connect concentration distributions, recurrence statistics, temporal memory, fluctuation amplification and local dynamical resilience. Using China as a primary case study, we show that the FMMR framework reproduces the observed statistical organization of PM$_{2.5}$ across pollution regimes, including distributional scaling, exceedance recurrence, increasing memory and reduced recovery under elevated pollution levels. We further apply the framework to multiple countries and populated, emission-influenced regions to examine how PM$_{2.5}$ resilience varies across pollution environments and evolves under recent air-quality improvements. By combining resilience diagnostics, recurrence statistics, meteorological analyses and population-relevant regional comparisons, we identify persistence risks that concentration metrics alone can miss and provide a quantitative basis for resilience-oriented air-quality assessment.

\section*{RESULTS}

\noindent\textbf{A unified multiplicative backbone of PM$_{2.5}$ variability}

\begin{figure}[!htbp]
	\centering
	\includegraphics[pagebox=cropbox,width=0.96\linewidth]{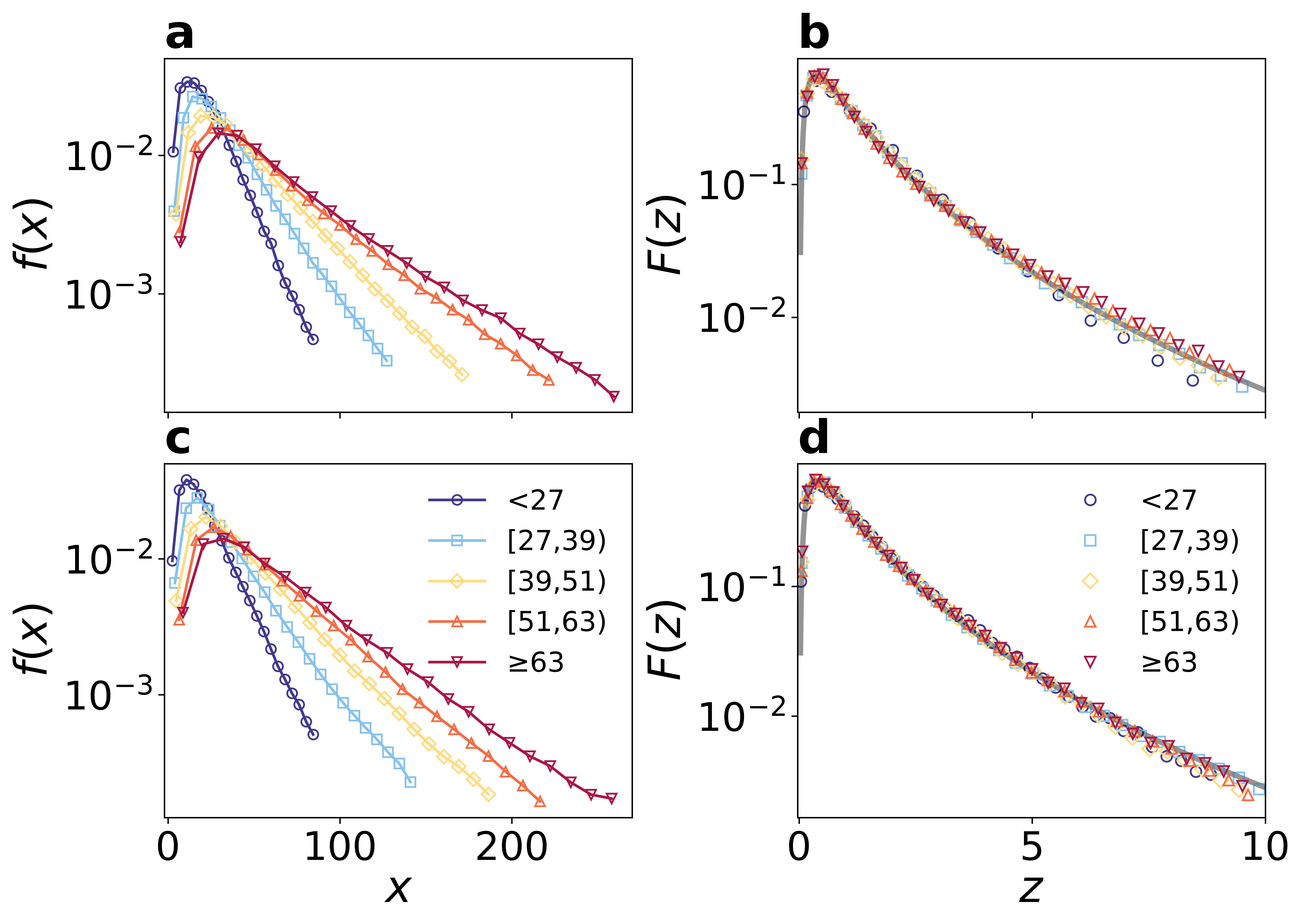}
	\caption{\textbf{Scaling of PM$_{2.5}$ concentration distributions in observations and FMMR simulations for China, revealing a universal multiplicative structure.} \textbf{a}, PDFs of PM$_{2.5}$ concentrations ($x$) for representative regions across China, grouped into five bins by regional mean concentration (approximately 15–75~$\mu$g~m$^{-3}$): $<27$, $27$–$39$, $39$–$51$, $51$–$63$, and $\geq 63$~$\mu$g~m$^{-3}$. \textbf{b}, Rescaled PDFs showing collapse onto a universal curve. \textbf{c}, \textbf{d}, Corresponding FMMR simulations for \textbf{a} and \textbf{b}, respectively. In \textbf{b} and \textbf{d}, the gray line denotes the universal scaling form given by Eq.~(\ref{eq:1}).}
	\label{11}
\end{figure}

\noindent Fig.~\ref{11}(a) shows the probability density functions (PDFs) of PM$_{2.5}$ concentrations across representative regions in China with distinct mean pollution levels over the study period (spatial locations are shown in Supplementary Fig.~1). Despite large differences in mean concentration, the distributions share a common structure: all PDFs are right-skewed, and the upper tail becomes progressively broader in more polluted regions. 

To quantify this behavior, we compare nine candidate distributions (Supplementary Fig.~2 and Supplementary Tables~1 and~2). Among these, the lognormal distribution provides the best overall fit across regions spanning different pollution regimes, indicating that PM$_{2.5}$ fluctuations are governed by a common multiplicative statistical structure even as pollution intensifies.

To further test whether this structure is universal, we introduce a normalized variable
$
Z=(X/x_{0})^{1/\sigma_Y},
$
where $x_0=\exp(\mu_Y)$ and $\mu_Y$ and $\sigma_Y$ are the mean and standard deviation of $Y=\ln X$, with $X$ denoting pollutant concentration. Under this transformation, PDFs from regions with widely different pollution levels collapse onto a single master curve (Fig.~\ref{11}(b)), consistent with the universal form
\begin{equation}
F(z)=\frac{1}{z\sqrt{2\pi}}\,\mathrm{e}^{-(\ln z)^2/2}.
\label{eq:1}
\end{equation}
This collapse reveals a unified lognormal backbone underlying PM$_{2.5}$ fluctuations across pollution regimes.

To explain this shared structure, we introduce a finite-memory multiplicative reversion (FMMR) process. In this reduced framework, pollutant concentration evolves relative to a characteristic baseline level $x_0$, representing the combined influence of emissions, meteorology, chemistry, transport and removal:
\begin{equation}
X_t=x_0\left(\frac{X_{t-1}}{x_0}\right)^\alpha \mathrm{e}^{\sigma \Delta W_t},
\label{eq:2}
\end{equation}
where $\Delta W_t$ denotes an increment of a Wiener process with amplitude $\sigma$. The parameter $\alpha$ is an effective memory parameter, governing how strongly current concentrations depend on recent pollution history, while $\sigma$ represents multiplicative stochastic forcing. When $\alpha$ is small, concentrations rapidly relax toward the baseline level in the reduced model. As $\alpha$ approaches unity, relaxation slows and temporal persistence strengthens, allowing pollution episodes to accumulate under conditions that may include weak ventilation, stable boundary layers, emissions variability and reduced removal. In logarithmic space, the FMMR process reduces to a first-order autoregressive process, implying a stationary Gaussian distribution for $\ln X$ and therefore a lognormal distribution for $X$ (see Methods).

Simulations based on the FMMR process reproduce the observed distributional patterns across pollution regimes (Fig.~\ref{11}(c,d); see Methods for details of simulations). In particular, the model captures both the right-skewed concentration PDFs and their collapse after normalization, indicating that a parsimonious finite-memory multiplicative representation can describe the common statistical backbone of PM$_{2.5}$ variability.

\noindent\textbf{Persistence and clustering of high pollution episodes}

\noindent We next examine how the temporal organization of high pollution events varies across pollution regimes. A pollution event is defined as a continuous interval during which PM$_{2.5}$ concentration exceeds a prescribed threshold $q$ within the 50th–90th percentile range of the overall distribution, and the waiting time as the interval between two consecutive exceedance events (Supplementary Fig.~3). To reduce the influence of seasonal variability, waiting times longer than 30 days are excluded.

\begin{figure}[!htbp]
	\centering
	\includegraphics[pagebox=cropbox,width=0.96\linewidth]{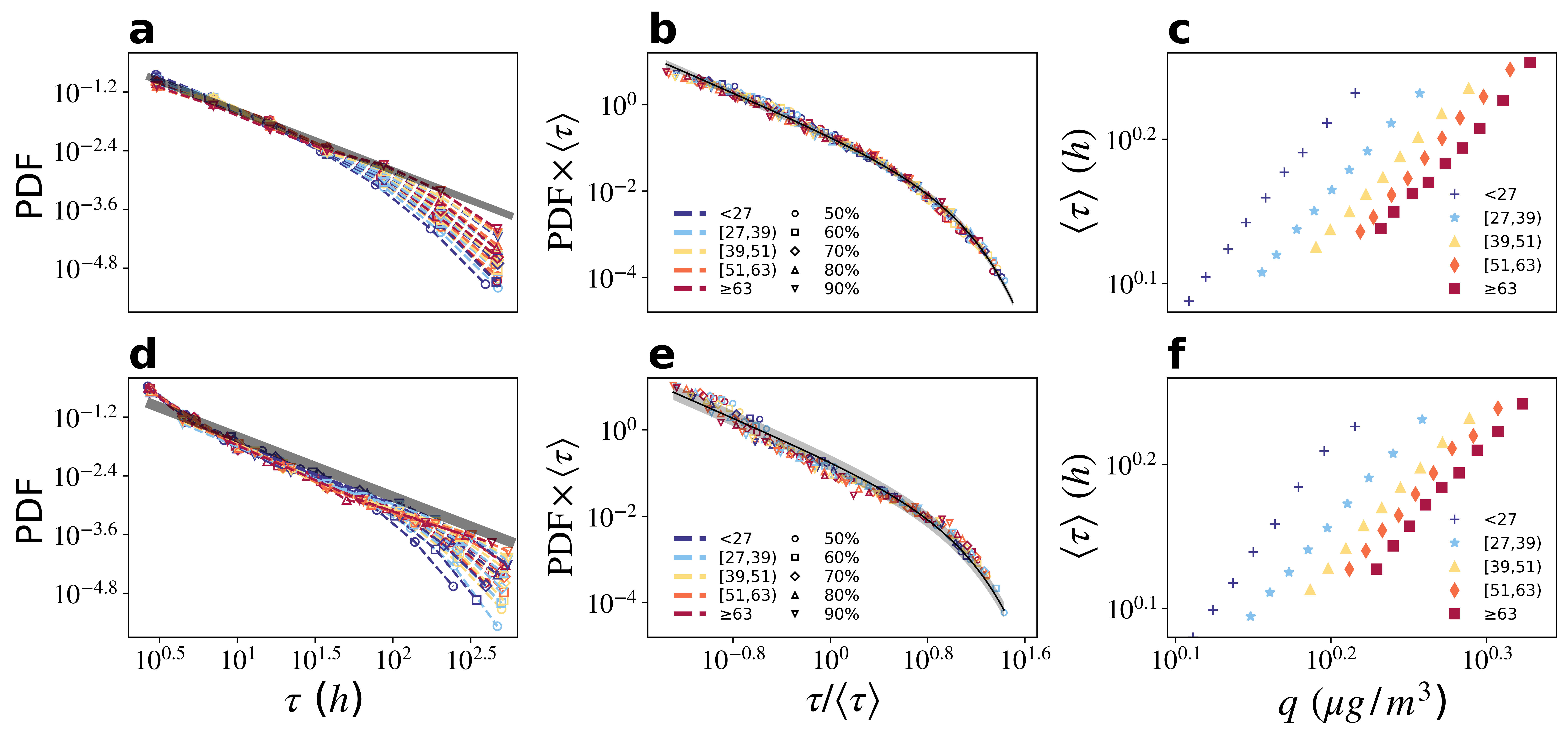}
		\caption{\textbf{Scaling of waiting-time distributions of PM$_{2.5}$ in observations and FMMR simulations for China, revealing common temporal organization of pollution events.} \textbf{a}, PDFs of waiting times for PM$_{2.5}$ across intervals of mean concentration; the black solid line indicates a reference algebraic slope of $-1.22$. \textbf{b}, PDFs rescaled by the mean waiting time $\langle \tau \rangle$, showing collapse onto a common curve; the solid black line denotes the fitted form given by Eq.~(\ref{eq:7}). \textbf{c}, Power-law relation between the mean waiting time $\langle \tau \rangle$ and the event threshold $q$. \textbf{d}--\textbf{f}, Corresponding FMMR simulations for \textbf{a}--\textbf{c}, respectively. In \textbf{e}, the solid black curve denotes the same fitted function and parameter values as in \textbf{b}. }
	\label{22}
\end{figure}

Across all concentration regimes, the waiting-time distributions exhibit a robust two-part structure consisting of finite-range algebraic scaling followed by an exponential cutoff at longer intervals (Fig.~\ref{22}(a)). After rescaling by the mean waiting time $\langle \tau \rangle$, the distributions collapse onto a single master curve (Fig.~\ref{22}(b)), well described by
\begin{equation}
  P(\tilde{\tau}) = C\,\tilde{\tau}^{-\gamma}\exp(-\kappa \tilde{\tau}),
\label{eq:7}
\end{equation}
where $\tilde{\tau}=\tau/\langle\tau\rangle$, $C$ is a normalization constant, $\gamma$ is the short-time scaling exponent, and $\kappa$ is the dimensionless cutoff rate. For the composite fit we obtain $C\simeq0.2$, $\gamma\simeq1.22$, and $\kappa\simeq0.15$. A comparison with exponential, pure power-law, lognormal and truncated-power-law candidate models confirms that this truncated-power-law form best captures the scaled waiting-time density (Supplementary Fig.~4). This collapse indicates that high PM$_{2.5}$ events share a common recurrence structure across regions despite large differences in background pollution levels, a pattern also observed in other natural hazard systems such as earthquakes, convective rainfall, and solar flares~\cite{16,20,21}.

Figure~\ref{22}(c) further reveals a power-law relation between the mean waiting time and the threshold $q$, such that
$
\langle \tau \rangle \sim q^{\beta},
$
or equivalently the event frequency $\lambda_q \sim 1/\langle \tau \rangle$ follows the inverse scaling. This threshold-scaling relation provides a compact recurrence indicator whose concentration dependence is examined together with memory, variance and autocorrelation below.

After renormalizing the $\langle \tau \rangle$–$q$ relation by the characteristic concentration $x_{0}$, the scaled relation preserves the recurrence organization across pollution regimes (Supplementary Fig.~5a,b). In relatively clean regions, the mean waiting time increases steeply with threshold, whereas in more polluted regions this dependence becomes weaker, indicating that severe exceedances become less separated in time. Seasonal concentration distributions and waiting-time statistics show the same organization, and the recurrence patterns remain robust to percentile-threshold ranges, log-space stationarization and waiting-time cutoffs (Supplementary Figs.~6--12).

Simulations based on the FMMR process reproduce these temporal signatures. The simulated waiting-time distributions display the same finite-range algebraic scaling with an exponential cutoff (Fig.~\ref{22}(d)), collapse onto the same rescaled form (Fig.~\ref{22}(e)) and recover the threshold scaling of the mean waiting time (Fig.~\ref{22}(f)). This indicates that multiplicative memory can account for the clustered recurrence of high-pollution episodes.

\noindent\textbf{Critical signatures strengthen under elevated PM$_{2.5}$}

\begin{figure}[!htbp]
\centering
\includegraphics[pagebox=cropbox,width=0.90\linewidth]{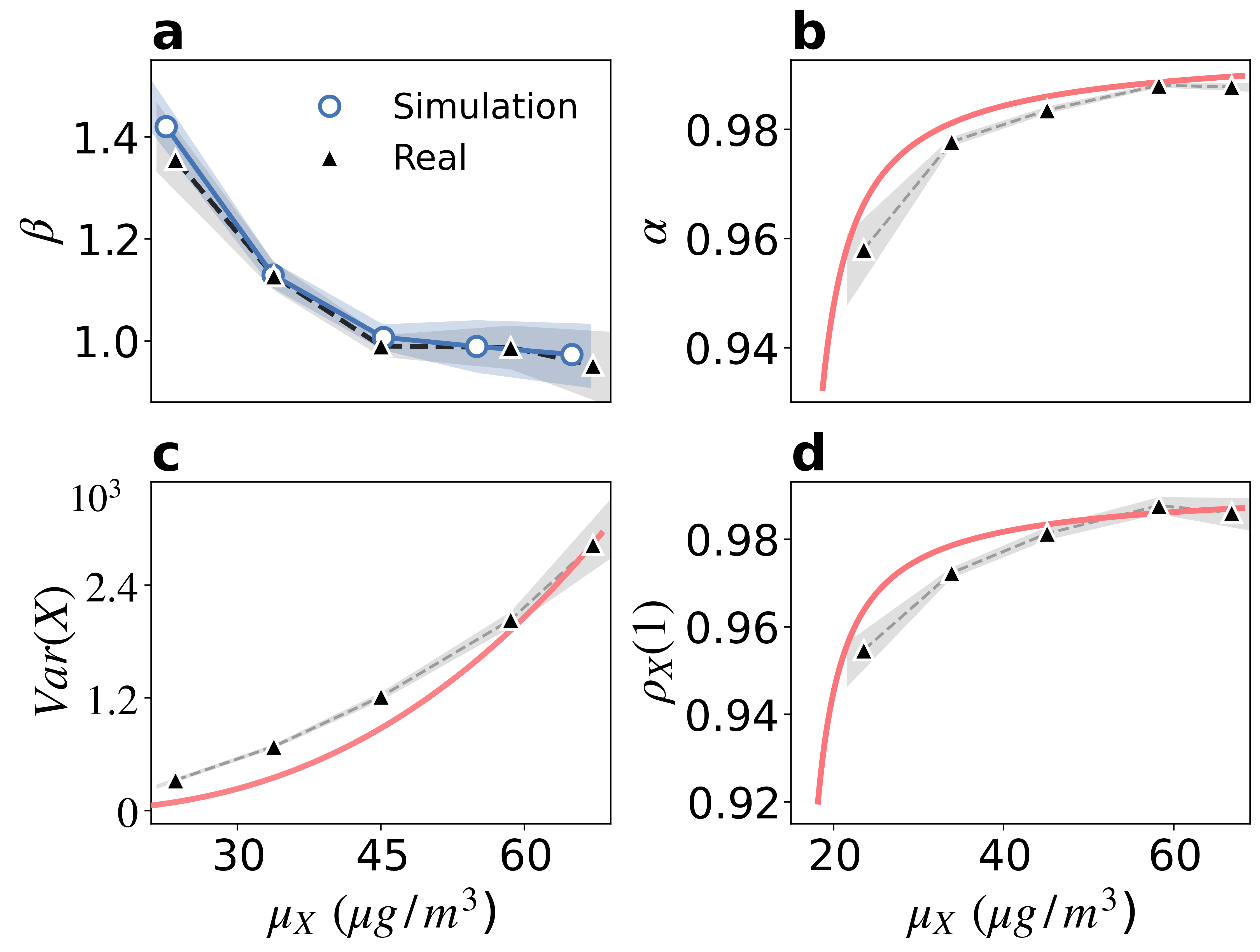}
\caption{\textbf{Observed and FMMR-simulated variations of scaling and dynamical indicators with pollution level for China.}
\textbf{a}, Power-law exponent $\beta$ as a function of the interval-mean concentration $\mu_X$ for observations and FMMR simulations.
The exponent $\beta$ is obtained from the power-law fits between the mean waiting time $\langle \tau \rangle$ and the event threshold $q$ shown in Fig.~\ref{22}c,f for observations and FMMR simulations, respectively.
\textbf{b}, Interval-mean memory coefficient $\alpha$, plotted against $\mu_X$.
\textbf{c}, Interval-mean variance $\mathrm{Var}(X)$, plotted against $\mu_X$.
\textbf{d}, Interval-mean AR(1) coefficient $\rho_X(1)$, plotted against $\mu_X$.
Shaded bands represent $\pm 1$ standard error estimated from a 365-day block-based resampling procedure; see Methods for details.
Red lines in \textbf{b}--\textbf{d} show analytical predictions from Eqs.~(\ref{eq:11})--(\ref{eq:13}). The analytical relation for \textbf{a}, described by Eq.~(\ref{eq:15}), is provided in Supplementary Fig.~5d.}
\label{33}
\end{figure}

\noindent We next combine the recurrence indicator with fitted memory, variance and autocorrelation to test whether critical signatures strengthen under elevated PM$_{2.5}$. Figure~\ref{33}(a) shows that the threshold-scaling exponent decreases with mean pollution level in both observations and FMMR simulations, indicating that high-pollution events become increasingly clustered and less distinguishable across thresholds in their temporal statistics. Within the FMMR framework, this behavior is controlled by the memory parameter $\alpha$, which sets the rate at which pollutant concentrations relax toward the characteristic level $x_0$. When $\alpha$ is small, concentrations rapidly revert to the baseline state. As $\alpha$ increases toward unity, relaxation slows and fluctuations become increasingly persistent.

Empirically, the fitted memory parameter increases systematically with the mean pollution level (Fig.~\ref{33}(b)). This dependence is captured by a semi-empirical relation linking $\alpha$ to the mean concentration $\mu_X$ (see Eq.~(\ref{eq:11}) in Methods). The analytical curve reproduces the observed trend that higher pollution levels correspond to stronger system memory and slower relaxation. Notably, the relation also implies a saturation effect: as pollution increases, $\alpha$ approaches unity and becomes progressively less sensitive to further increases in concentration.

A direct consequence of this strengthening memory is the amplification of concentration variance (see Eq.~(\ref{eq:12}) in Methods). In the FMMR process, multiplicative perturbations are amplified by longer memory, producing a super-linear variance--mean relationship (Fig.~\ref{33}(c)). This mechanism reproduces the variance inflation observed in highly polluted regimes and contributes to the broadening of concentration distributions.

Temporal persistence provides an additional indicator of reduced recovery. The AR(1) coefficient increases monotonically with the mean PM$_{2.5}$ concentration and approaches unity as pollution intensifies (Fig.~\ref{33}(d); see Eq.~(\ref{eq:13}) in Methods). Similar behavior is observed in hourly/daily AR(1) diagnostics and daily-scale parameter relations (Supplementary Figs.~13 and~14). In dynamical systems, such behavior is consistent with critical slowing down, where perturbations decay increasingly slowly. Here it forms one component of the broader criticality inferred from the co-occurrence of memory growth, variance amplification, upper-tail broadening and clustered exceedance events. These concentration-dependent trends persist after log-space stationarization and under alternative spatial aggregation choices (Supplementary Figs.~15--17), indicating that they are not artefacts of the mean seasonal cycle, a linear trend or the specific grid size.

A physical pathway for this enhanced persistence is boundary-layer suppression~\cite{Li2017ABLReview,Miao2019PBLPM25Review}. Severe pollution episodes are typically associated with shallow and stable planetary boundary layers that suppress turbulent mixing and reduce atmospheric ventilation~\cite{22,add1,CAP2024STOTEN}. Consistent with this mechanism, observational analysis shows a negative relationship between boundary-layer height (BLH) and the daily AR(1) coefficient (Supplementary Fig.~18). This relationship remains negative after controlling for wind speed, relative humidity, temperature and total precipitation through partial-correlation analysis, whereas the corresponding partial correlations for these other meteorological variables are weaker and less spatially consistent (Supplementary Fig.~19). Seasonal analysis further shows that the negative BLH--AR(1) relationship is retained in both warm and cold seasons (Supplementary Fig.~20), and BLH decreases across increasingly polluted concentration intervals (Supplementary Fig.~21). These results identify BLH suppression as an important meteorological contributor to the enhanced persistence measured by $\alpha$, alongside synoptic transport, humidity-dependent aerosol growth, wet removal, emissions variability and aerosol chemistry.

Analytical approximations derived from the FMMR framework further link the threshold-scaling exponent to the effective memory parameter (see Eq.~(\ref{eq:15}) in Methods), explaining why denser event recurrence emerges as the system enters a higher-persistence regime.

Finally, we test whether the same dynamical organization appears beyond China. PM$_{2.5}$ observations from India display similar statistical patterns and scaling relations (Supplementary Figs.~22--24), indicating that comparable persistence organization appears in another heavily polluted environment. In contrast, PM$_{2.5}$ dynamics in the United States show substantially lower persistence, consistent with lower pollution levels (Supplementary Fig.~25).

\noindent\textbf{Contrasting resilience states across world regions}

\begin{figure}[!htbp]
\centering
\includegraphics[pagebox=cropbox,width=0.6\linewidth]{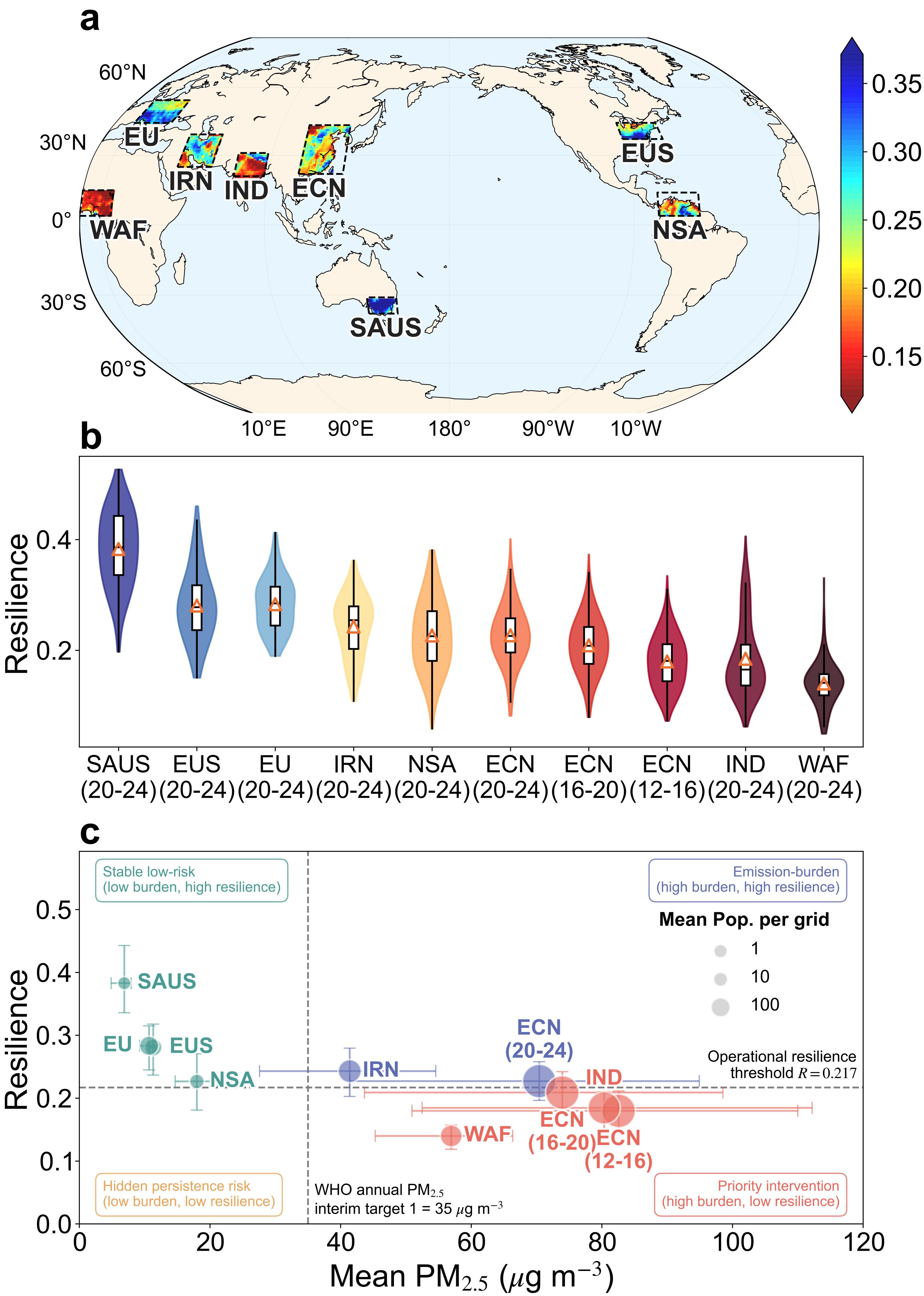}
\caption{\textbf{Resilience of PM$_{2.5}$ pollution systems across populated and emission-influenced world regions.}
\textbf{a}, CAMS-based dynamical resilience ($1-\alpha$) for 2020--2024; dashed boxes mark representative regions (selection and definitions in Supplementary Note~4). \textbf{b}, Regional resilience distributions for populated land cells. Boxes show 25th--75th percentiles, lines show medians and triangles show means; ECN is shown for three periods. \textbf{c}, Pollution burden--resilience state space. Points are regional means; error bars show 25th--75th percentiles; size denotes mean population per grid cell. Dashed lines mark the 35~$\mu$g~m$^{-3}$ burden reference and the CAMS-based operational resilience reference level (0.217; median of 2020--2024 regional median resilience values).}
\label{44}
\end{figure}

\noindent Having established concentration-dependent critical signatures in station observations, we next ask whether the associated resilience metric, defined below as $1-\alpha$, separates pollution burden from recovery capacity across world regions. Figure~\ref{44} presents a reanalysis-based comparison of PM$_{2.5}$ pollution resilience across populated and emission-influenced regions using CAMS fields and a WorldPop population mask (see Methods and Supplementary Note~4). Within the FMMR framework, resilience is quantified as $1-\alpha$, representing the relaxation rate of localized perturbations. Low resilience indicates stronger memory and weaker restoring capacity, whereas high resilience reflects faster relaxation toward the baseline concentration $x_0$. The CAMS gridded framework provides a common basis for comparing regional resilience states across different pollution environments.

Figure~\ref{44}a shows the spatial distribution of resilience during 2020--2024. Regions such as southeastern Australia (SAUS), the eastern United States (EUS) and Europe (EU) generally exhibit higher resilience, indicating faster recovery from PM$_{2.5}$ perturbations. By contrast, India (IND) and West Africa (WAF) show lower resilience, indicating stronger memory and greater persistence risk. Iran (IRN), northern South America (NSA) and eastern China (ECN) show more heterogeneous patterns.

Figure~\ref{44}b further compares the regional distributions of resilience using violin-box plots. The temporal evolution of ECN is especially notable: from 2012--2016 to 2020--2024, the distribution shifts upward from a lower-resilience regime toward a higher-resilience state. This change is consistent with the substantial reductions in PM$_{2.5}$ concentrations achieved under sustained air-quality control measures in China. However, the 2020--2024 ECN distribution remains below the high-resilience levels observed in SAUS, EUS and EU, indicating that parts of eastern China still retain appreciable persistence risk. This CAMS-based recovery signal over eastern China is further supported by CNEMC station observations, which show an upward shift in resilience from 2016--2020 to 2021--2024 (Supplementary Fig.~26).

To separate pollution burden from dynamical recovery, we further project the representative regions into a pollution burden--resilience state space (Fig.~\ref{44}c). The horizontal axis represents the mean PM$_{2.5}$ burden, whereas the vertical axis represents the mean resilience. This representation shows that regions with similar pollution levels can differ substantially in recovery capacity. The 0.217 reference level is defined as the median of the 2020--2024 regional median resilience values and is used here as a CAMS-based empirical divider. ECN moves toward lower PM$_{2.5}$ burden and higher resilience over time, approaching and then exceeding this reference level. This joint shift shows that recent air-quality improvement in ECN is associated not only with lower mean concentration, but also with faster dynamical recovery. In contrast, IND is in the high-burden, low-resilience part of the state space, combining elevated PM$_{2.5}$ with weak recovery and a large exposed population. WAF occupies a different but important state: although its mean PM$_{2.5}$ burden is slightly lower than that of IND, its resilience remains low, indicating persistence risk under future emission growth or unfavourable meteorological conditions.

The regional spreads in Fig.~\ref{44}c also reveal substantial within-region heterogeneity. ECN, IND and IRN exhibit broad horizontal error bars, indicating large spatial differences in PM$_{2.5}$ burden across populated land grid cells. By contrast, SAUS, EUS and EU occupy the low-burden, high-resilience part of the state space and show comparatively narrower regional spreads. These contrasts indicate that resilience provides information complementary to concentration-based metrics, because it captures the persistence and recovery properties of regional pollution systems.

We next test whether this separation between pollution burden and recovery carries forward in time. Using 2015--2019 as the baseline and 2020--2024 as the evaluation period, we compare subsequent exceedance-event duration against baseline mean PM$_{2.5}$ and baseline resilience. The duration--concentration relationship is non-monotonic and relatively diffuse (Supplementary Fig.~27a). By contrast, event duration decreases more systematically with baseline resilience (Supplementary Fig.~27b), indicating that low-resilience regions tend to experience longer pollution episodes in the subsequent period.

To ask whether the same baseline information also organizes broader atmospheric changes, we compare period-to-period responses in total precipitation, cloud ice water and cloud liquid water. These responses show a more coherent nonlinear organization when grouped by baseline resilience than when grouped by baseline mean PM$_{2.5}$ concentration (Supplementary Fig.~28). The strongest changes occur around this CAMS-based reference level (0.217): below this value, precipitation and cloud-water changes are generally suppressed, whereas above it they show a tendency toward partial recovery. This organization may reflect differences in ventilation, aerosol loading, wet removal and aerosol--cloud--precipitation interactions~\cite{Yang2017PM25Meteorology,Tao2012AerosolCloud,Fan2016AerosolCloud}. Together, the event-duration and atmospheric-response analyses indicate that resilience is not merely a restatement of pollution level, but a dynamical diagnostic linked to future pollution persistence and broader atmospheric responses.

\section*{DISCUSSION}

This study provides a stochastic framework for diagnosing criticality in atmospheric PM$_{2.5}$ dynamics by linking multiplicative variability, event recurrence, temporal memory, variance amplification and local dynamical resilience. Across station observations and reanalysis data, elevated pollution regimes show a coherent set of critical signatures: broader upper tails, stronger memory, higher autocorrelation, clustered exceedance events and reduced resilience. Together, these patterns reveal criticality in PM$_{2.5}$ pollution systems, with critical slowing down appearing as one manifestation of reduced restoring capacity.

Within the FMMR framework, the memory parameter $\alpha$ connects statistical structure to dynamical response. As $\alpha$ increases, relaxation slows, temporal dependence strengthens and perturbations decay more slowly. The negative relationship between boundary-layer height and temporal persistence supports weak ventilation and boundary-layer stability as important contributors to this increased memory, while synoptic transport, humidity-dependent aerosol growth, wet removal, emissions variability and aerosol chemistry may also contribute. Accordingly, $\alpha$ should be interpreted as a net persistence diagnostic of the reduced stochastic process, not as a uniquely identified physical cause.

The regional comparison translates this dynamical diagnosis into a sustainability-relevant risk dimension. India and West Africa occupy lower-resilience parts of the CAMS-based state space, whereas Europe and the eastern United States exhibit higher resilience consistent with lower pollution levels and faster recovery. Eastern China has shifted toward higher resilience during recent air-quality improvements, although parts of the region still retain persistence risk. These contrasts show that PM$_{2.5}$ burden and recovery capacity can diverge, so concentration reductions may improve both exposure burden and dynamical recovery.

The FMMR framework is phenomenological rather than a first-principles model of emissions, chemistry and meteorology, and the resilience metric used here is a dynamical recovery rate rather than the broader socio-environmental meanings of resilience. The observed critical signatures indicate reduced restoring capacity and increasingly persistent dynamics under elevated PM$_{2.5}$, but they do not by themselves establish a phase transition or imply that a tipping point must exist. With this scope made explicit, persistence and recovery diagnostics provide a complementary basis for air-quality assessment and sustainability-oriented pollution management, helping identify regions where pollution risk is sustained by slow atmospheric recovery rather than by concentration burden alone.

\section*{METHODS}

\noindent\textbf{The FMMR process}

\noindent The FMMR process is specified in discrete time by Eq.~(\ref{eq:2}). Taking logarithms transforms the multiplicative dynamics into an additive first-order autoregressive process,
\begin{eqnarray}
  Y_t &=& \ln X_t, \nonumber\\
      &=& \alpha\, Y_{t-1} + \sigma\, \Delta W_t + \eta, \label{eq:m2}
\end{eqnarray}
where $\eta=(1-\alpha)\ln x_{0}$. Here $x_0$ is the characteristic concentration level, $\alpha$ is the effective memory parameter, and $\sigma$ is the amplitude of multiplicative stochastic forcing. For $|\alpha|<1$, the stationary distribution of $Y$ is Gaussian,
\begin{eqnarray}
  Y &\sim& \mathcal{N}\!\left(\frac{\eta}{1-\alpha},\, \frac{\sigma^{2}}{1-\alpha^{2}}\right)
  =
  \mathcal{N}\!\left(\ln x_{0},\, \frac{\sigma^{2}}{1-\alpha^{2}}\right), \label{eq:4}
\end{eqnarray}
implying that $X = \mathrm{e}^{Y}$ follows a lognormal distribution. As $\alpha$ increases toward unity, the effective memory length increases and relaxation toward $x_0$ slows. In the limiting case $\alpha=0$, the process has no temporal memory in log space, whereas $\alpha=1$ removes mean reversion.

\noindent\textbf{Linking pollution level to effective memory}

\noindent For a Gaussian variable $Y$, the mean of $X=\exp(Y)$ is $\mu_X = \exp(\mu_Y + \frac{1}{2}\sigma_Y^{2})$. Using Eq.~(\ref{eq:4}), we obtain
\begin{eqnarray}
  \ln \mu_X &=& \frac{\eta}{1-\alpha} + \frac{\sigma^{2}}{2\,(1-\alpha^{2})}\,. \label{eq:8}
\end{eqnarray}
In the high-persistence limit $\alpha\to1$, $(1-\alpha^{2})\approx 2(1-\alpha)$, so Eq.~(\ref{eq:8}) becomes
\begin{eqnarray}
  \ln \mu_X &\approx& \frac{\,\eta+\sigma_{0}\,}{1-\alpha}\,, \label{eq:9}
\end{eqnarray}
where $\sigma_{0}=\sigma^{2}/4$. The characteristic level $x_0$ can co-vary with $\alpha$ because emissions, chemical production, removal and meteorological ventilation jointly shape both the baseline concentration and the persistence of pollution episodes. We therefore write $x_0=x_0(\alpha)$ locally. On the finite left neighbourhood $\alpha\in(1-\varepsilon,1)$, assuming that $\ln x_0$ is twice continuously differentiable, a one-sided Taylor expansion gives, uniformly in $\alpha$,
\begingroup\small
\begin{equation}
\begin{aligned}
\eta(\alpha) &= (1-\alpha)\,\ln x_{0}(\alpha) \\
             &= -\ln x_{0}(1)\,\alpha + \bigl[\ln x_{0}(1)+\Delta_{\varepsilon}\bigr],
\end{aligned}
\label{eq:10}
\end{equation}
\endgroup
where $|\Delta_{\varepsilon}|\le K\varepsilon^{2}$. This expansion explains why the drift term is expected to be approximately linear in $\alpha$ near the high-persistence regime. Consistent with this expectation, the station-based PM$_{2.5}$ series show an empirical linear approximation $\eta(\alpha)=\theta_{0}-\theta_{1}\alpha$ (Supplementary Fig.~5c), with $\theta_{0}\approx 2.713$, $\theta_{1}\approx 2.70$ and $\sigma_{0}\approx 0.0025$. Combining $\eta(\alpha)=(1-\alpha)\ln x_0(\alpha)$ with this empirical relation gives
\begin{equation}
  \ln x_{0}(\alpha)=\theta_{1}+\frac{\theta_{0}-\theta_{1}}{1-\alpha}.
\label{eq:10_1}
\end{equation}
Thus, if $\theta_{0}=\theta_{1}$, the characteristic level is locally independent of $\alpha$; the observed positive difference $\theta_{0}-\theta_{1}$ indicates a weak positive coupling between the baseline concentration level and effective memory. Substituting $\eta(\alpha)=\theta_{0}-\theta_{1}\alpha$ into Eq.~(\ref{eq:9}) gives
\begin{eqnarray}
  \alpha &\approx& 1-\frac{\theta_{0}-\theta_{1}+\sigma_{0}}{\ln \mu_X-\theta_{1}}\,. \label{eq:11}
\end{eqnarray}
Equation~(\ref{eq:11}) gives the analytical curve relating effective memory to mean pollution level in Fig.~\ref{33}b.

The corresponding variance and lag-1 autocorrelation predictions follow from lognormal moment identities. Using the variance identity derived in Supplementary Note~9,
\begin{equation}
  \mathrm{Var}(X) = \mathrm{E}(X^{2}) - \mathrm{E}(X)^{2} = \mu_X^{2}\Bigl[\exp\!\Bigl(\tfrac{4\sigma_{0}}{1-\alpha^{2}}\Bigr)-1\Bigr],
  \label{eq:12}
\end{equation}
and substituting $\alpha(\mu_X)$ from Eq.~(\ref{eq:11}) yields the analytical variance--concentration relation in Fig.~\ref{33}c. Similarly, the closed-form lag-1 autocorrelation of $X$, derived in Supplementary Note~8, is
\begin{equation}
  \rho_{X}(1)=\frac{\exp\!\left(\dfrac{4\alpha\,\sigma_{0}}{1-\alpha^{2}}\right)-1}
                  {\exp\!\left(\dfrac{4\sigma_{0}}{1-\alpha^{2}}\right)-1}.
  \label{eq:13}
\end{equation}
Together with Eq.~(\ref{eq:11}), Eq.~(\ref{eq:13}) yields the AR(1) prediction in Fig.~\ref{33}d.

For exceedance events, Kac's lemma~\cite{25} relates the mean waiting time to the inverse exceedance probability. Under the high-persistence upper-tail approximation derived in Supplementary Note~10, the exponent $\beta$ in the $\langle\tau\rangle$--$q_X$ scaling satisfies
\begin{eqnarray}
  \beta &\approx& \frac{1}{2\,\sigma_Y^{2}} \;=\; \frac{1-\alpha^{2}}{8\,\sigma_{0}}\,. \label{eq:15}
\end{eqnarray}
Combining Eqs.~(\ref{eq:11}) and~(\ref{eq:15}) gives the analytical curve in Supplementary Fig.~5d. Low-$\mu_X$ discrepancies are expected because the approximation is asymptotic in the high-persistence limit.

\noindent\textbf{System resilience}

\noindent In dynamical systems, resilience is commonly defined as the relaxation rate of localized perturbations. For the FMMR process, we define local dynamical resilience from the deterministic mean-reversion component corresponding to the discrete model,
\begin{equation}
  f(X) = -(1-\alpha)X \ln\left(\frac{X}{x_0}\right).
\end{equation}

The fixed point is obtained by solving $f(X^\star)=0$, which yields $X^\star = x_0$ for $X>0$. Linearizing the drift around this point gives the local eigenvalue governing perturbation dynamics:
\begin{equation}
  \xi = f'(x_0) = \left. -(1-\alpha) \left[ \ln\left(\frac{X}{x_0}\right) + 1 \right] \right|_{X=x_0} = -(1-\alpha).
\end{equation}

For $0 \le \alpha < 1$, we have $\xi < 0$, indicating local stability. The system resilience is therefore defined as the corresponding decay rate of perturbations:
\begin{equation}
  \text{Resilience} = -\xi = 1-\alpha.
\end{equation}

This result shows that the memory parameter $\alpha$ directly controls the restoring strength within the FMMR model. As $\alpha \to 1$, the resilience $1-\alpha$ approaches zero, indicating weaker recovery and critical slowing down. In contrast, when $\alpha = 0$, resilience reaches its maximum value, corresponding to rapid relaxation toward the baseline level $x_0$ and the absence of temporal memory in the reduced process.

\noindent\textbf{Parameter estimation and simulation}

\noindent FMMR parameters are estimated in log space for each grid-cell time series. For a length-$N$ series $y_t=\ln X_t$ with mean $\mu_Y$ and standard deviation $\sigma_Y$, the AR(1) memory parameter is estimated as
\begin{equation}
  \alpha =
  \frac{\sum_{t=2}^{N} (y_t - \mu_Y)(y_{t-1} - \mu_Y)}
       {\sum_{t=2}^{N} (y_{t-1} - \mu_Y)^2}.
\end{equation}
The innovation scale and characteristic concentration are then estimated as
\begin{equation}
  \sigma = \sqrt{1-\alpha^{2}}\,\sigma_Y,
\end{equation}
\begin{equation}
  x_0 = e^{\mu_Y}.
\end{equation}
The corresponding drift term is $\eta=(1-\alpha)\mu_Y$, and $\sigma_0=\sigma^2/4$.
These estimates are used both to quantify memory and resilience in observations and reanalysis fields and to generate FMMR simulations. For the simulations shown in Figs.~\ref{11} and~\ref{22}, we use the fitted parameters for each grid cell and generate hourly concentration series of length $22\,000$ steps, discarding the first $2\,000$ steps to remove the initial adjustment period.

\noindent\textbf{Data sources and preprocessing}

\noindent To ensure comparability across regions, we treat all three PM$_{2.5}$ station datasets within a common framework. Together, these datasets provide the observational backbone of our analysis: CNEMC and EPA AirNow underpin the China--U.S. comparison, and CPCB supplies an independent third-region validation over India. All three datasets are processed with a harmonized quality-control and gridding workflow, and all concentrations are expressed in $\mu\mathrm{g\,m}^{-3}$, ensuring that the subsequent metrics and regional comparisons are comparable and reproducible. 

We use hourly PM$_{2.5}$ observations from the China National Environmental Monitoring Center (CNEMC) for 13 May 2014–31 December 2021 over eastern China (105$^\circ$–127$^\circ$E, 22$^\circ$–50$^\circ$N), obtained via the “China Air Quality Historical Data” (\url{https://quotsoft.net/air/}). Stations with more than 15\% missing data are discarded. For each hour, records with PM$_{2.5}>500~\mu\mathrm{g\,m}^{-3}$ are treated as outliers and removed, and the remaining stations are averaged on a $2^\circ\times2^\circ$ lat–lon grid. Grid cells with fewer than two valid stations at a given hour are excluded. Residual gaps are filled by linear interpolation in time, yielding a continuous gridded dataset.

For comparison with China, we use hourly PM$_{2.5}$ observations from the U.S. EPA AirNow network (\url{https://www.airnow.gov/}) over the contiguous United States (25$^\circ$–49$^\circ$N, 66$^\circ$–123$^\circ$W) for 1 January 2015–31 December 2021~\cite{Lee2024CAMSvsAirNow}. The same preprocessing is applied as for CNEMC: removal of stations with more than 15\% missing data, exclusion of hourly values above 500~$\mu\mathrm{g\,m}^{-3}$, aggregation to a $2^\circ\times2^\circ$ grid requiring at least two valid stations per cell, and linear interpolation in time to fill isolated gaps.

Given the relative sparsity of the Indian monitoring network, we use hourly PM$_{2.5}$ observations from the Central Pollution Control Board (CPCB; main portal \url{https://cpcb.nic.in/}) for 1 January 2022–31 December 2023. To ensure consistency, we adopt the same quality-control and gridding strategy, with a slightly coarser spatial resolution: outliers above 500~$\mu\mathrm{g\,m}^{-3}$ are removed; station data are aggregated to a $3^\circ\times3^\circ$ grid with at least two valid stations per cell; and remaining temporal gaps are filled by linear interpolation.

Additional preprocessing sensitivity tests are summarized in Supplementary Note~1. Briefly, we quantified the distribution of continuous missing-record lengths before interpolation (Supplementary Fig.~29) and used the Tracking Air Pollution in China (TAP) gridded PM$_{2.5}$ product as an independent consistency check without applying additional temporal interpolation in our workflow (Supplementary Fig.~30).

To complement the station-based analyses and to compare resilience across major world regions within a unified gridded framework, we additionally use PM$_{2.5}$ fields from the Copernicus Atmosphere Monitoring Service (CAMS) global reanalysis (EAC4)~\cite{Sowden2024CAMSEval}. The dataset is provided on a regular $0.75^\circ \times 0.75^\circ$ latitude--longitude grid and is analyzed at 3-hourly resolution, yielding eight time steps per day. Since the native CAMS PM$_{2.5}$ mass concentration is provided in SI units, it is converted to $\mu\mathrm{g\,m}^{-3}$ for analysis.

To support the physical-attribution and out-of-sample atmospheric-response analyses, we additionally use meteorological fields from the ERA5 reanalysis~\cite{Hersbach2020ERA5}, including boundary-layer height, wind components, temperature, dew-point temperature, total precipitation, cloud ice water and cloud liquid water.

In the spatial and regional analyses shown in Fig.~\ref{44}, $\alpha$ is estimated from the log-transformed CAMS time series at each retained land grid cell, and resilience is calculated as $1-\alpha$. Land cells are retained using a 2020 WorldPop population mask. The operational resilience reference level, $R_{\mathrm{ref}}=0.217$, is defined as the median across the 2020--2024 regional median resilience values of the representative regions and is used within this CAMS-based analysis as a common empirical divider rather than as a physical critical threshold. The mask and representative regional boxes are described in Supplementary Note~4 and shown in Supplementary Fig.~31. We use CAMS as a unified reanalysis-based framework for broad interregional comparison, while recognizing that regional differences in reanalysis skill, emissions inventories, aerosol representation and observational constraints may affect the resilience patterns.

\noindent\textbf{Out-of-sample evaluation}

\noindent To evaluate whether the resilience metric contains information beyond the baseline pollution level, we performed an out-of-sample evaluation with 2015--2019 as the reference period and 2020--2024 as the evaluation period. Baseline mean PM$_{2.5}$, resilience and a fixed local high-percentile threshold were estimated from the reference period and used to organize subsequent exceedance-event duration and atmospheric responses (Supplementary Figs.~27 and~28). Detailed definitions are provided in Supplementary Note~5.

\noindent\textbf{Multivariate meteorological attribution analysis}

\noindent To assess whether the observed BLH--AR(1) relationship remains robust after accounting for other meteorological covariates, we performed a multivariate attribution analysis over eastern China. Daily PM$_{2.5}$ persistence was paired with ERA5 BLH, wind speed, relative humidity, temperature and total precipitation, and both Pearson correlations and residual-based partial correlations were calculated. The residual-based procedure and uncertainty estimates are described in Supplementary Note~6.

\noindent\textbf{Uncertainty estimation}

\noindent The shaded bands in Fig.~\ref{33} were estimated using a 365-day block-based sampling scheme that combines temporal-block variability with spatial variability across grid cells. Details are given in Supplementary Note~7.

\section*{ACKNOWLEDGEMENTS}
This work was supported by the National Natural Science Foundation of China (Grant No. 12305044) and the National Key Research and Development Program of China (Grant No. 2023YFE0109000). SH and YA acknowledge support from the Council for Higher Education of Israel through the climate science call, under the project ``Integrating Climate Dynamics, Clouds, and Extreme Events through Teleconnections in Climate Networks''. We also acknowledge the data resources provided by the China National Environmental Monitoring Center (CNEMC), the U.S. Environmental Protection Agency (EPA AirNow), and the Central Pollution Control Board (CPCB) of India.


\end{document}